\documentclass[conference]{IEEEtran}
\usepackage{cite}
\usepackage{amsmath,amssymb,amsfonts}
\usepackage{algorithmic}
\usepackage{graphicx}
\usepackage{textcomp}
\usepackage{xcolor}
\usepackage{url}
\usepackage{listings}
\usepackage{comment}
\usepackage{cuted}     % for strip
\usepackage{capt-of}   % for \captionof

\usepackage{graphicx}
\usepackage{tabularx}
\usepackage{booktabs}
\usepackage[caption=false, font=footnotesize]{subfig}
\usepackage{float}

\def\BibTeX{{\rm B\kern-.05em{\sc i\kern-.025em b}\kern-.08em
    T\kern-.1667em\lower.7ex\hbox{E}\kern-.125emX}}
\begin{document}

\title{SeedAIchemy: LLM-Driven Seed Corpus Generation for Fuzzing}

\author{
% \thanks{Identify applicable funding agency here. If none, delete this.}

\IEEEauthorblockN{Aidan Wen\IEEEauthorrefmark{1}, Norah A. Alzahrani\IEEEauthorrefmark{3}\IEEEauthorrefmark{4}, Jingzhi Jiang\IEEEauthorrefmark{1}\IEEEauthorrefmark{4}, Andrew Joe\IEEEauthorrefmark{1}\IEEEauthorrefmark{4}, Karen Shieh\IEEEauthorrefmark{1}\IEEEauthorrefmark{4}, Andy Zhang\IEEEauthorrefmark{1}\IEEEauthorrefmark{4}, \\
Basel Alomair\IEEEauthorrefmark{2}, David Wagner\IEEEauthorrefmark{1}}

\IEEEauthorblockA{\IEEEauthorrefmark{1}University of California, Berkeley}
\IEEEauthorblockA{\IEEEauthorrefmark{2}King Abdulaziz City for Science and Technology}
\IEEEauthorblockA{\IEEEauthorrefmark{3}Humain, Saudi Arabia}
\IEEEauthorblockA{\IEEEauthorrefmark{4}These authors contributed equally to this research}
}

\maketitle

\begin{abstract}
% Fuzz testing is a widely used method for improving software security. Fuzzing is most effective when using a seed corpus with a diverse set of inputs. Constructing such a corpus can be difficult for developers who are newly adopting fuzz testing or do not have a strong security background. 

We introduce SeedAIchemy, an automated LLM-driven corpus generation tool that makes it easier for developers to implement fuzzing effectively. SeedAIchemy consists of five modules which implement different approaches at collecting publicly available files from the internet. Four of the five modules use large language model (LLM) workflows to construct search terms designed to maximize corpus quality. Corpora generated by SeedAIchemy perform significantly better than a naive corpus and similarly to a manually-curated corpus on a diverse range of target programs and libraries. 
\end{abstract}

\begin{IEEEkeywords}
fuzzing, large language models, computer security, software testing 
\end{IEEEkeywords}

\section{Introduction}
Fuzz testing is a widely used method for improving software security.
One of the attractions of fuzz testing is that it is relatively easy to adopt.
However, one road bump with adopting fuzz testing is that, for best effectiveness,
developers must provide a corpus of seed files.
Ideally, these seed files would include many tricky cases and difficult inputs,
and would ensure good branch coverage of the targets.
Constructing such a corpus can be difficult for developers who are newly adopting
fuzz testing or do not have a strong security background.

One approach that some developers take when initially adopting fuzz testing
is to use a naive corpus: e.g., a corpus containing only a single trivial file.
Such a naive corpus is easy for developers to collect and makes adoption easy,
but it also makes fuzzing less effective.
In contrast, collecting a high-quality corpus can require considerable effort
and some sophistication and knowledge of computer security.
This leaves developers with a tradeoff between ease of adoption and effectiveness.

In this paper, we seek to support developers and make it easier
to adopt fuzz testing by automating the process of collecting a corpus.
We introduce SeedAIchemy, a tool that automates the entire process of corpus collection,
so adopting fuzz testing can be easy and effective, even for developers
who are not deeply knowledgeable about security.
SeedAIchemy builds a corpus by collecting seed files from publicly available files on the internet.
It guides its searches using a large language model  (LLM) workflow that constructs
search terms customized for the target being tested and aims to find files useful for fuzzing.

We show in our evaluation that SeedAIchemy is effective on a wide range of file types.
In our experiments on the Magma benchmark, SeedAIchemy generates a corpus
that makes fuzzing significantly more effective than a naive corpus:
fuzzing with AFL++ for 24 hours triggers $1.7\times$ as many unique bugs
with a SeedAIchemy-generated corpus compared to a naive corpus on average
(naive corpus: 20 bugs; SeedAIchemy corpus: 34 bugs).
Moreover, the SeedAIchemy corpus approaches the high-quality
manually-curated corpus provided with the Magma benchmark:
fuzzing with the SeedAIchemy finds 97\% of the bugs revealed by the Magma corpus on average
(SeedAIchemy: 34 bugs, Magma corpus: 35 bugs).
Of course, SeedAIchemy requires no human effort, whereas collecting the
Magma corpus required considerable human curation.

Overall, we believe this demonstrates that LLM agents can be helpful
for reducing barriers to adoption of fuzz testing, and specifically,
for building a useful fuzzing corpus.
By automatically generating seed corpora with similar quality to manually-curated corpora,
% While manual curation by a security expert is still slightly more
% effective than SeedAIchemy's automated method, 
SeedAIchemy provides an
excellent starting point for developers who are not security experts or
do not have resources for extended manual curation.
In the rest of this paper, we explain the techniques used to build
SeedAIchemy and we evaluate its effectiveness with detailed experiments.

\section{Problem Statement}\label{sec:problem_statement}
Fuzz testing (fuzzing) is a well-established and widely used technique for discovering bugs and security vulnerabilities in software \cite{manes2019art}. Fuzzing software works by taking in a target intended to be tested, also known as the \textit{fuzz target}, and running it repeatedly with a series of random inputs, with the goal of triggering unexpected behaviors \cite{manes2019art, klees2018evaluating}. Over the past decade, fuzzing has uncovered tens of thousands of vulnerabilities in critical real-world software. For example, Google’s OSS-Fuzz, a fuzzing platform which contains over a thousand open-source projects, has uncovered more than 10,000 vulnerabilities and 36,000 bugs as of August 2023 \cite{ossfuzz2023}. As another example, the Syzbot fuzzing project has discovered over 5570 bugs in the Linux Kernel from 2018-2024, leading to over 4604 bug fixes \cite{bursey2024syzretrospector}. 

Fuzzers generate random inputs by modifying a set of initial valid inputs known as the \textit{seed corpus} \cite{manes2019art}. A diverse and representative corpus can help fuzzers reach deeper executions and find subtle bugs \cite{jurczyk2016effective, herrera2021seed}. However, due to the difficulty and time involved, many projects in practice pay little attention to the construction of the seed corpus and use either a single seed file or none at all \cite{klees2018evaluating, herrera2021seed}. 

Recent progress in large language models (LLMs) presents a new opportunity to make fuzzing easier to adopt. LLMs can be leveraged to produce creative, context-guided search queries to systematically mine large-scale information sources such as web search engines. Even if LLMs hallucinate or generate poor quality search queries, the worst case is that the generated seed files provide no additional coverage and are discarded. 
% LLMs have demonstrated surprising ability in other tasks \cite{gpt4-leetcode-2024}. 
We show that LLMs can be similarly applied to seed corpus generation in place of human computer security experts.

We have developed SeedAIchemy, an easy-to-use corpus building tool that automatically generates seed files using LLM-generated search queries. 
SeedAIchemy generates corpora by combining the results of five modules which implement different strategies for finding seed files.
By reducing the time and effort needed to construct a seed corpus, we hope more developers will use fuzzing in their development process and do so with higher effectiveness.
 
% To show that our tool improves fuzzing effectiveness on real-world, large-scale software systems, we investigate whether corpora generated by SeedAIchemy can improve bug discovery and code coverage, two fundamental performance metrics  \cite{klees2018evaluating}. In our evaluation, bug discovery is quantified using two measures: bugs reached and bugs triggered. A bug is ``reached" if the vulnerable code regions are executed during fuzzing and ``triggered" if the conditions for revealing the bug are satisfied. Code coverage is quantified using bitmap coverage as reported by AFL++, a state-of-the-art coverage-guided fuzzing tool \cite{aflplusplus-docs}. We run fuzzing on the Magma benchmark and compare the results of corpora generated by our tool against naive and manually constructed corpora. We also compare corpora generated by our tool to G$^2$FUZZ \cite{Zhang2025G2FUZZ}, an existing LLM-based corpus generation tool.

\section{Design}
As discussed in Section \ref{sec:problem_statement}, a well constructed corpus can greatly improve the effectiveness of fuzzing  \cite{klees2018evaluating}. Our goal in developing SeedAIchemy was to provide a way to automatically generate seed corpora with sufficient size, diversity, and relevance for effective fuzzing on a wide range of fuzz targets. 
% This section describes our approach, beginning with the general architecture of the tool and followed by more details for each submodule. 

\subsection{General approach}
\begin{figure}[tb]
    \centering
    \includegraphics[width=\linewidth]{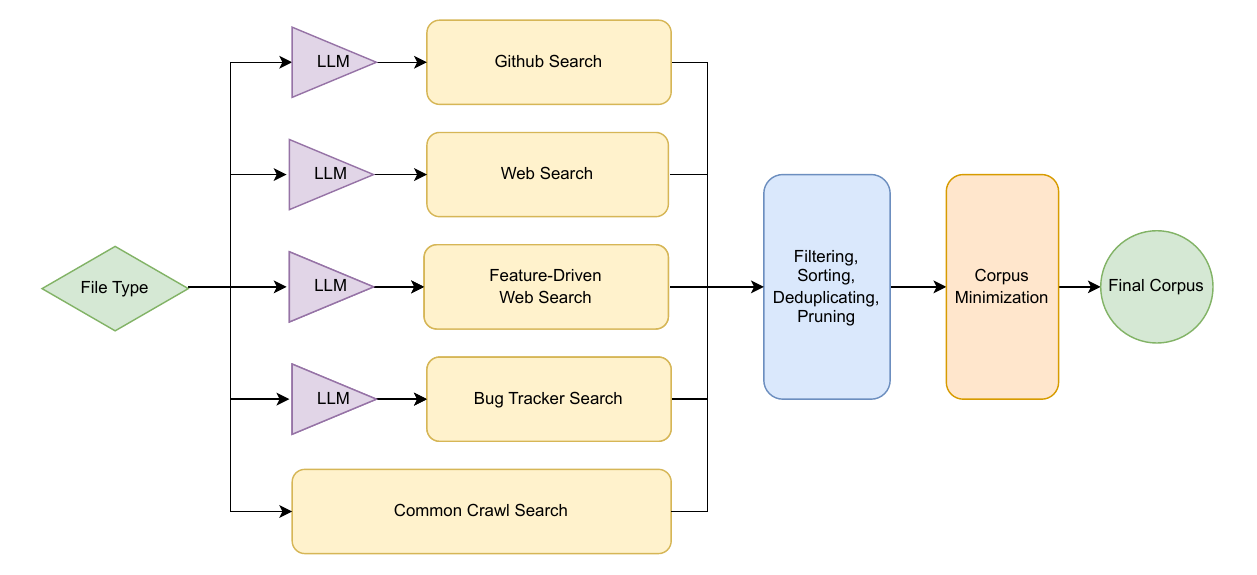}
    \caption{Architecture of SeedAIchemy. SeedAIchemy combines corpora from each submodule, then applies minimization techniques to reduce the size of the final corpus.}
    \label{fig:general_architecture}
\end{figure}

\begin{figure}[tb]
    \centering
    \includegraphics[width=\linewidth]{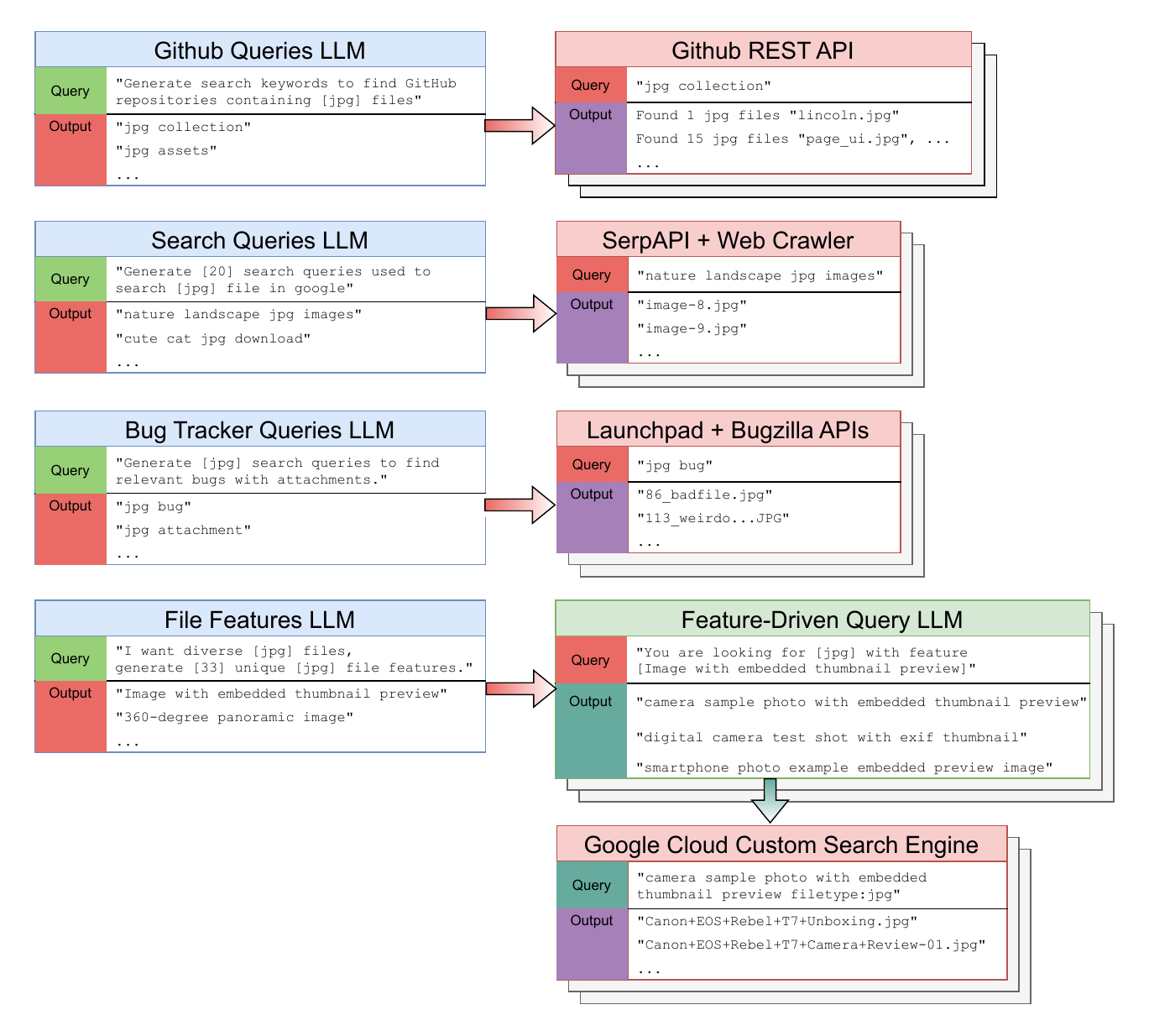}
    \caption{Example queries and outputs of a single run of SeedAIchemy for the JPG datatype. To preserve space, LLM prompts are simplified versions of the real ones used in SeedAIchemy. Only modules that used an LLM to generate search queries are shown.}
    \label{fig:example_process}
\end{figure}

To produce a large and varied seed corpus, SeedAIchemy delegates seed file generation to five modules, each using a different seed gathering technique, as outlined in \figurename~\ref{fig:general_architecture} and \ref{fig:example_process}. To focus on seed files that are relevant to the fuzz target, the user provides a file extension: for instance, the user might specify ``pdf'' if they are fuzzing a PDF viewer. The techniques detailed in Sections B--F are then executed in parallel. Each module generates a subcorpus of seed files with the target file extension.
The resulting subcorpora are combined and deduplicated, and files larger than the specified maximum file size (default: 1 MB) are filtered out to improve fuzzing efficiency. 


Finally, SeedAIchemy selects a subset of seed files for the final corpus.
If there are more than 40,000 candidates, it selects the smallest files first and ensures a balance across modules.
This corpus could be used directly for fuzzing, but for better effectiveness, we recommend using a corpus minimization tool such as afl-cmin.
We provide further details in Section G.

Each module generates search terms using an LLM, then queries a particular search engine.  The GitHub and feature-driven web search modules use GPT-4.1; web search and bug tracker search use GPT-4o.  Common Crawl search does not use an LLM; it deterministically searches archived web content for files with the specified file extension. Each module runs for at most 1 hour.

%To further refine the final corpus, a maximum of 40,000 seed files is imposed. For file types that appear frequently on GitHub (e.g. PDF), the number of potential seeds remains excessively large even after deduplication and size filtering. To this end, our design first includes files from other submodules and then includes the GitHub files, selecting for the smallest ones first. This ensures that the final corpus is representative of a diverse range of sources, thereby maximizing the quality \cite{aflplusplus-docs} \cite{herrera2020corpusdistillationeffectivefuzzing}.

%The resulting seed files are a functional corpus that can be used immediately for fuzzing, but to improve performance, we recommend using a corpus minimization tool such as afl-cmin or an equivalent. We detail this process and the rationale for it in section G. 

%The specific LLM used in each module is shown in Table \ref{tab:llm_for_each_module}. Note that \textit{Common Crawl Search} does not use a LLM to generate seach queries; rather, it deterministically searches over archived web content using the given file extension. 

\begin{comment}
\begin{table}[htbp]
    \centering
    \begin{tabular}{ll}
        \toprule
        \textbf{Module} & \textbf{LLM} \\
        \midrule
        GitHub Search & GPT-4.1 \\
        Web search & GPT-4o \\
        Feature-driven Web Search & GPT-4.1 \\
        Bug Tracker Search & GPT-4o \\
        Common Crawl & none \\
        \bottomrule
    \end{tabular}
    \caption{LLM used in each module}
    \label{tab:llm_for_each_module}
\end{table}
\end{comment}

\subsection{GitHub Search}

The GitHub Search module uses LLM-generated search queries to find seed files on GitHub. 
% We ask an LLM to generate a set of diverse search keywords and then search GitHub with those search terms, finding the repository that each match appears in. Files matching the desired file type are extracted from each repository and added to the output corpus.
We prompt GPT-4.1 to generate a set of 50 diverse search queries that look for repositories that are broadly used or directly relevant to software testing. For each search query, we find the top 10 matches and shallow-clone the associated repository to minimize bandwidth and storage usage. Cloning repositories rather than downloading individual files reduces the number of API calls and allows users to stay within GitHub’s rate limits. For each cloned file, we check the file extension and magic number to determine whether the file is the correct format. Files satisfying either condition are retained.

\subsection{Web search}
The Web Search module looks for seed files using a general search on the internet. We ask GPT-4o to generate a mix of 20 distinct queries that capture the characteristics of the desired file type (e.g. ``small PNG clipart" or ``PDF ebooks") as well as queries that explicitly look for existing seed files (e.g. ``fuzzing seed JPG files"). We then search Google with those queries, download each search result, and follow links recursively to collect all files matching the specified extension. We use a Scrapy spider \cite{scrapinghub2020scrapy} to process multiple webpages in parallel, which speeds up the scraping process and allows us to find more files within the 1 hour timeout period. 

\subsection{Feature-driven Web Search}
Another way to increase the diversity of a seed corpus is to use seed files that contain different features or structures. For example, a PDF file can contain images, Javascript code, or forms, and a high quality corpus should include as many of these features as possible. This module aims to find seed files that cover a wide range of available features for a given file type.

The Feature-driven Web Search module uses GPT-4.1 to identify features relevant to a file type and generate corresponding search queries. We use one-shot prompting to instruct GPT-4.1 to produce 33 distinct feature descriptors (e.g., “PNG with a visible watermark”, “PNG containing embedded GPS/location metadata”). We then ask it to expand each feature keyword into three complete search queries (e.g., “sample stock photo png with watermark download”, “technical diagram png export with dpi metadata”). We then search Google with each query, fetch the top 10 results, and crawl them recursively with a Scrapy spider without depth restriction \cite{scrapinghub2020scrapy}. Downloaded files are included in the final corpus if they match the correct file extension or magic number \cite{qti3e_extensionsjson}.

\subsection{Bug Tracker Search}

Bug trackers provide a good source of diverse, high-quality input files \cite{acharya2025bugsrepocomprehensivecurateddataset}. Many bug reports include a reproducing test case as an attachment alongside the bug description and commit. When manually crafting seed corpora, security experts often use files that have triggered bugs in the past \cite{rebert2014optimizing}. This is the motivation behind the Bug Tracker Search module, which searches popular bug tracker and reporting sites for files that have triggered bugs in the past.  

The Bug Tracker Search module works by prompting an LLM to generate search queries to find bug reports that are relevant to the target file type and have associated attachments. It uses the generated queries to search Ubuntu Launchpad and Red Hat Bugzilla using the launchpadlib API and the Bugzilla REST API. The search results are then crawled to collect attachments that match the target file type by checking for matches with the file extension or the magic number. We focus on these two bug trackers because of their popularity, extensive use in open-source ecosystems, and comprehensive API support.

\subsection{Common Crawl Search}

Common Crawl \cite{commoncrawl2025} is an open source project started in 2007 that crawls the web and public archive data.  Each crawl retrieves tens to hundreds of terabytes of web data, making it one of the largest openly available projects of web crawl data. The Common Crawl Search module searches the Common Crawl archive hosted on Amazon Web Services \cite{awsopendata2025}, specifically, the \textit{CC-MAIN-2025-08} crawl, by performing an exact match search on the \textit{content\_mime\_type} metadata. The \textit{CC-MAIN-2025-08} archive contains files with over 2,000 different types, providing a rich resource for fuzzing seeds of various formats.

\subsection{Corpus minimization}
SeedAIchemy can find up to several hundred thousand files depending on the prevalence of the file type. However, fuzzing performance is best when the seed corpus is small because smaller corpora increase execution speed and allow fuzzers to quickly discard uninteresting inputs \cite{herrera2021seed}. One way to reduce the size of a corpus is to use a \text{minimization} tool, which finds the smallest subset of seed files that maintains the same code coverage as the original corpus \cite{herrera2020corpusdistillationeffectivefuzzing}. Thus, the final step of SeedAIchemy involves running the corpus through afl-cmin, which is a well regarded minimization tool that has been shown to improve fuzzing performance \cite{herrera2020corpusdistillationeffectivefuzzing, herrera2021seed}.

afl-cmin minimizes corpora with edge coverage as its metric for fuzzing performance \cite{afldocs}. Measuring coverage can be slow, so afl-cmin can be slow.  To ensure that afl-cmin completes in a reasonable amount of time, we discard all files larger than 1 MB and limit the output corpus to at most 40,000 files by preferentially selecting the 40,000 smallest files before minimization.

\subsection{Files without extensions}
SeedAIchemy allows users to search for files using a file description or phrase (e.g., "php\_serialize"). This feature works similarly to the file extension based corpus generation, with two modifications: LLMs generate search queries using a file description rather than a file extension, and file extension and magic number filtering are disabled. This feature allows SeedAIchemy to generate seed corpora for fuzz targets whose input files do not have any standard file extensions (e.g., binary blobs), which makes it useful for a wider range of targets. 

\section{Evaluation}

We evaluate SeedAIchemy on the Magma fuzzing benchmark.
Magma provides a high-quality manually-curated corpus for each target, mostly gathered from the original library repositories and other sources. 
We compare fuzzing with a SeedAIchemy-generated corpus to fuzzing with a naive corpus, the Magma corpus, and a corpus generated by G$^2$FUZZ, an existing LLM-based corpus generation tool \cite{Zhang2025G2FUZZ}. 

We follow robust evaluation and benchmarking principles proposed in prior work \cite{klees2018evaluating}, \cite{schloegel2024sok}, \cite{bohme2022reliability}. For example, Klees et al. suggest adhering to the following guidelines:
\begin{itemize}
    \item Run multiple fuzzing trials and evaluate results using statistical tests.
    \item Use fuzzing timeouts of at least 24 hours.
    \item Have a diverse set of benchmark programs.
    \item Test multiple seed corpora.
    \item Measure performance in terms of known bugs.
\end{itemize}
The first two conditions ensure that the evaluation procedure accounts for the randomness of fuzzing, the next two ensure that the proposed fuzzing technique is effective in a diverse range of fuzzing conditions, and the last condition ensures that the evaluation uses a concrete metric for fuzzing performance. Our evaluation follows these guidelines by carrying out the following:
\begin{itemize}
    \item Run 10 trials for each corpus and fuzz target pair. We compute a 95\% confidence interval for each aggregated statistic and compare results using Wilcoxon signed-rank tests.
    \item Use a fuzzing timeout of 24 hours.
    \item Test fuzz targets from the Magma benchmark.
    \item Compare SeedAIchemy to three corpora of differing quality.
    \item Measure performance in terms of bugs reached, bugs triggered, and code coverage.
\end{itemize}

\subsection{Experimental method} 
Our evaluation was conducted using AFL++, a widely used coverage-guided mutation-based greybox fuzzer \cite{aflplusplus-docs}. We measured the performance of AFL++ running for 24 hours on fuzz targets seeded with four different corpora as shown in Table \ref{tab:evaluation_corpora}:
\begin{itemize}
    \item \textbf{Naive}: a corpus consisting of a single file randomly sampled from the Magma benchmark corpus. This represents a corpus that a developer with minimal fuzzing experience could possibly use.
    \item \textbf{G$^2$FUZZ}: a corpus generated using the G$^2$FUZZ Input Generator Synthesis subpart, an LLM-based corpus generation tool proposed by Zhang et al. \cite{Zhang2025G2FUZZ}. G$^2$FUZZ generates seed files by asking an LLM to produce a Python script that generates seed files with the correct format. SeedAIchemy differs from this approach by asking the LLM to generate search queries rather than the file itself. 
    \item \textbf{SeedAIchemy}: our corpus generated with a maximum runtime of 1 hour, then minimized with afl-cmin using a timeout of 1 second per seed file.
    \item \textbf{Magma}: the corpus from the Magma benchmark \cite{Hazimeh:2020:Magma}, representing a corpus manually crafted by a domain expert with knowledge of the fuzz target. 
\end{itemize}

\begin{table}[tb]
    \centering
    \caption{Evaluation Corpora}
    \begin{tabularx}{\linewidth}{lX}
        \toprule
            \textbf{Corpus} & \textbf{Description} \\
        \midrule
            Naive & Randomly sampled file from the Magma corpus \\
            G$^2$FUZZ & LLM generated corpus \cite{Zhang2025G2FUZZ} \\
            SeedAIchemy & SeedAIchemy corpus minimized with afl-cmin \\
            Magma & Magma benchmark corpus \\
        \bottomrule        
    \end{tabularx}
    \label{tab:evaluation_corpora}
\end{table}

Each corpus was evaluated on fuzz targets from Magma v1.2 \cite{Hazimeh:2020:Magma}, a ground-truth fuzzing benchmark that forward-ports real bugs into modern versions of widely used software programs. Because Magma contains known bugs, it measures bug coverage without stack hashes or other bug deduplication techniques that are known to be unreliable \cite{klees2018evaluating}. Magma includes nine real-world targets spanning diverse domains and input formats shown in Table \ref{tab:magma_targets_and_file_types}.
It measures performance in terms of bugs reached and bugs triggered; a bug is ``reached" if the associated lines of code are executed and ``triggered" if the conditions for revealing the bug are satisfied. To support fuzzing with custom corpora, we ran our experiments on a fork of Magma commit \verb|75d1ae7| that contains our own modifications.

\begin{table}[tb]
    \caption{Magma Targets and Input File Types}
    \centering
    \begin{tabularx}{\linewidth}{l X l}
        \toprule
        \textbf{Libraries} & \textbf{Programs} & \textbf{File type(s)} \\
        \midrule
        libpng & libpng\_read\_fuzzer & .png \\
        libsndfile & sndfile\_fuzzer & audio files \\
        libtiff & tiff\_read\_rgba\_fuzzer & .tiff \\
         & tiffcp & .tiff \\
        libxml2 & libxml2\_xml\_read\_memory\_fuzzer & .xml \\
         & xmllint & .xml \\
        lua & lua & .lua \\
        openssl & asn1 & binary blobs \\
         & asn1parse & binary blobs \\
         & bignum & binary blobs \\
         & client & binary blobs \\
         & server & binary blobs \\
         & x509 & binary blobs \\
        php & exif & .jpeg, .tiff \\
         & json & .json \\
         & parser & .php \\
         & unserialize & serialized php \\
        poppler & pdf\_fuzzer & .pdf \\
         & pdfimages & .pdf \\
         & pdftoppm & .pdf \\
        sqlite3 & sqlite3\_fuzzer & .sql \\
        \bottomrule
    \end{tabularx}
    \label{tab:magma_targets_and_file_types}
\end{table}

For each fuzz target and corpus pair, we performed 10 independent fuzzing trials and evaluated performance using three metrics: (a) average number of bugs triggered, (b) average number of bugs reached, and (c) average code coverage as measured by AFL++ bitmaps. For each bug, we also measured the average time to reach and trigger that bug. Because LLMs produce nondeterministic output, we generated a new corpus for each SeedAIchemy fuzzing trial. To ensure consistency across trials, all experiments were run on a machine with two Intel(R) Xeon(R) Gold 5320 CPUs (26 physical cores each) and 504 GB of RAM, running Ubuntu 20.04.6 LTS.

\subsection{Cost analysis}
SeedAIchemy is inexpensive to deploy, costing only a few cents for each generated corpus. A 1-hour trial consumes approximately 0.5k input tokens and 0.7k output tokens on GPT-4.1, and consumes 12.5k input tokens and 2.3k output tokens on GPT-4o. This costs roughly \$0.06 based on pricing at the time of evaluation (GPT-4.1: \$2.00 per million uncached input tokens and \$8.00 per million output tokens; GPT-4o: \$2.50 per million uncached input tokens and \$10.00 per million output tokens) \cite{openai_pricing}. The cost of AWS Athena/S3 is harder to estimate as it depends on the input file extension. However, based on pricing at the time of evaluation \cite{awsathena} \cite{awss3}, generating a corpus costs only a few cents. SeedAIchemy is compatible with the free tiers of Google Search API and SerpAPI, and the cost of CPU and GPU usage, internet, and storage is negligible at this scale.

\section{Results}
The corpora produced by SeedAIchemy outperform naive and G$^2$FUZZ corpora on the Magma benchmark according to all three evaluation metrics. Compared to the Magma corpora, SeedAIchemy reaches 99\% of the bugs that Magma does, triggers 96\% as many bugs, and attains 91\% of Magma's code coverage (Table \ref{tab:results_summary_numerical}), showing that corpora produced by SeedAIchemy are of similar quality to a manually constructed corpus.

\begin{table}[tb]
    \caption{Average Bugs Reached, Bugs Triggered, and Normalized Code Coverage for Evaluated Corpora}
    \centering
    \begin{tabular}{lccc}
        \toprule
        \textbf{Corpus} & \textbf{Reached} & \textbf{Triggered} & \textbf{Coverage} \\
        \midrule
        Naive & 44.5 & 19.9 & 0.68 \\
        G$^2$FUZZ & 55.8 & 25.1 & 0.76\\
        SeedAIchemy & 75.9  & 33.9 & 0.91\\
        Magma & 76.7 & 35.3 & 1.00 \\
        \bottomrule
    \end{tabular}
    
    \label{tab:results_summary_numerical}
\end{table}

\subsection{Bug finding ability}

\begin{figure}[tbp]
    \centering
    \subfloat[Bugs Reached]{\includegraphics[width=\linewidth]{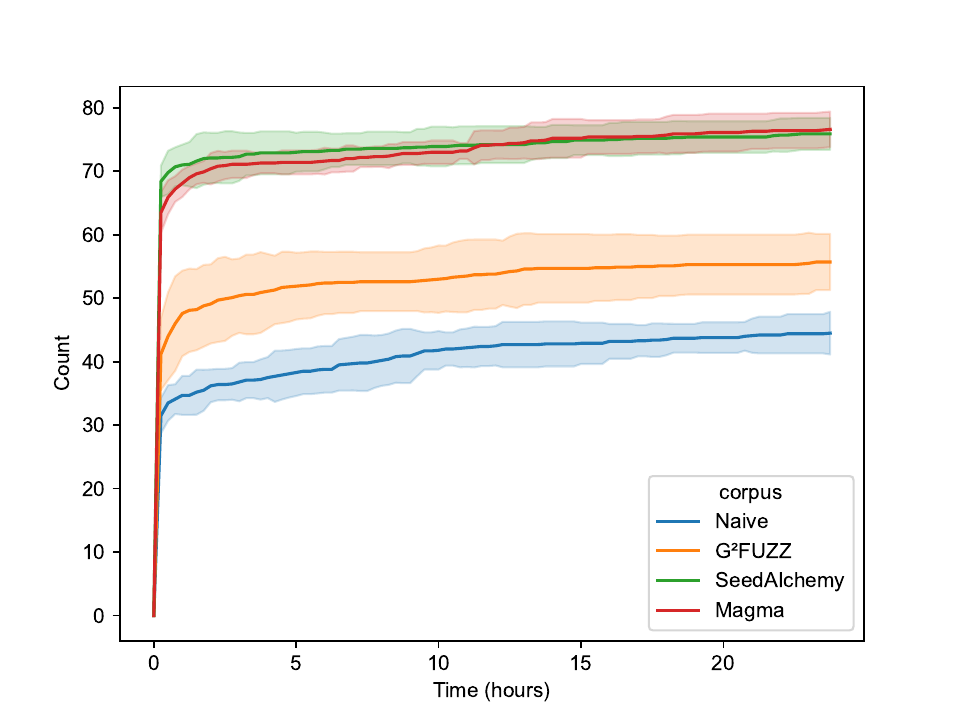}\label{fig:bugs_reached_over_time}}
    \hfil
    \subfloat[Bugs Triggered]{\includegraphics[width=\linewidth]{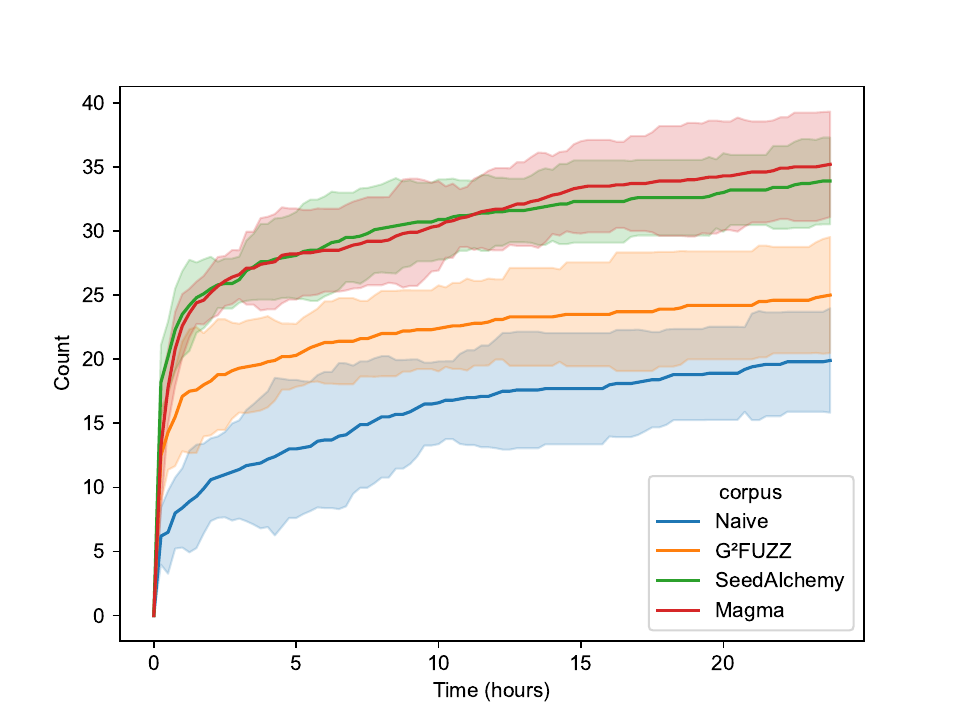}\label{fig:bugs_triggered_over_time}}
    \hfil
    \subfloat[Code Coverage (normalized)]{\includegraphics[width=\linewidth]{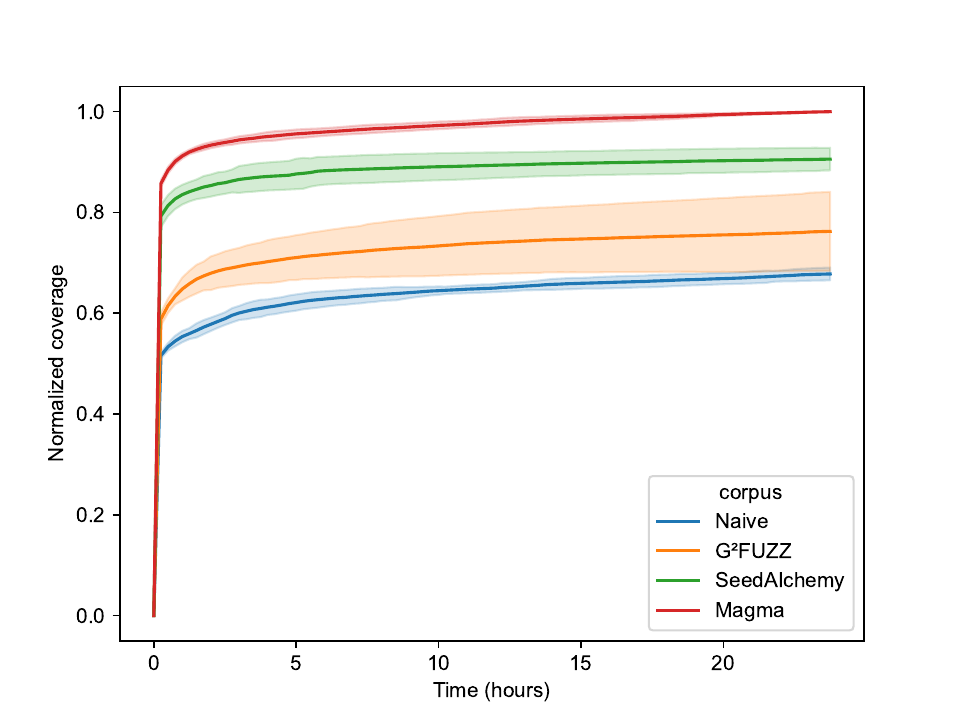}\label{fig:code_coverage_over_time}}

    \caption{Bugs reached, bugs triggered, and normalized code coverage averaged over 10 trials. Error bands show the 95\% confidence interval. Coverage is normalized by the 24-hour coverage of each Magma fuzz target.}
    \label{fig:magma_aggregate_lineplots}
\end{figure}

\figurename~\ref{fig:bugs_reached_over_time} and \ref{fig:bugs_triggered_over_time} show the total number of bugs reached and bugs triggered for each corpus averaged across 10 trials. SeedAIchemy reached and triggered $1.7\times$ as many bugs as a naive corpus. Also, it reached and triggered $1.4\times$ as many bugs as the G$^2$FUZZ corpus. It reached 99\% as many bugs as Magma and triggered 96\% as many bugs. Overall, after 24 hours of fuzzing, SeedAIchemy and Magma achieved similar numbers of bugs reached and bugs triggered while the Naive and G$^2$FUZZ corpora performed significantly worse. 

\begin{figure}[tbp]
    \centering
    \subfloat[Bugs Reached]{\includegraphics[width=\linewidth]{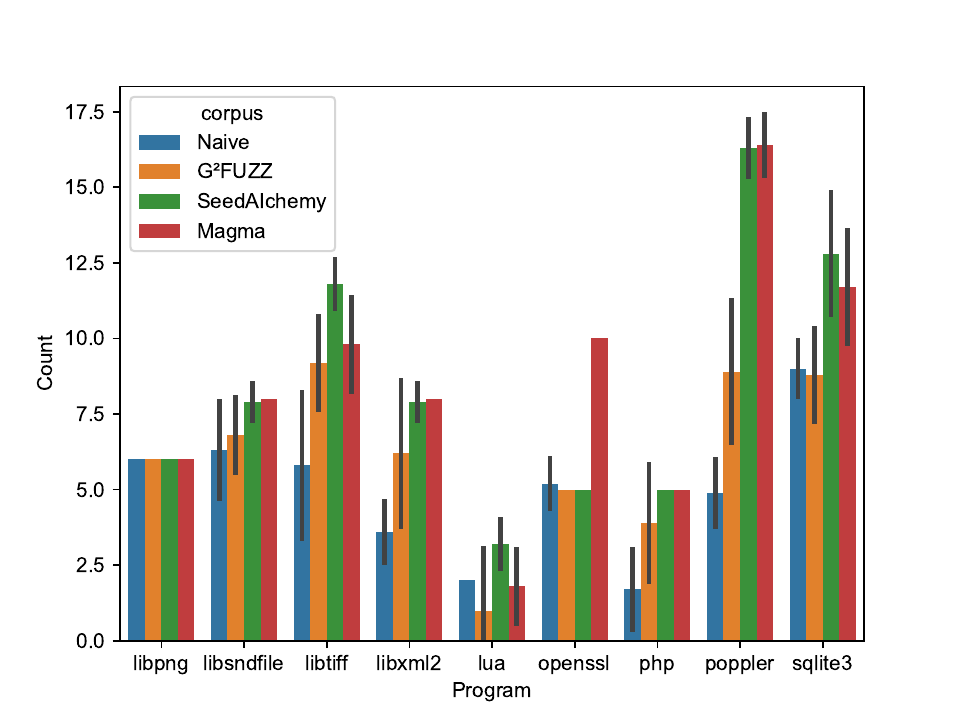}\label{fig:magma_reached_bar_chart}}
    \hfil
    \subfloat[Bugs Triggered]{\includegraphics[width=\linewidth]{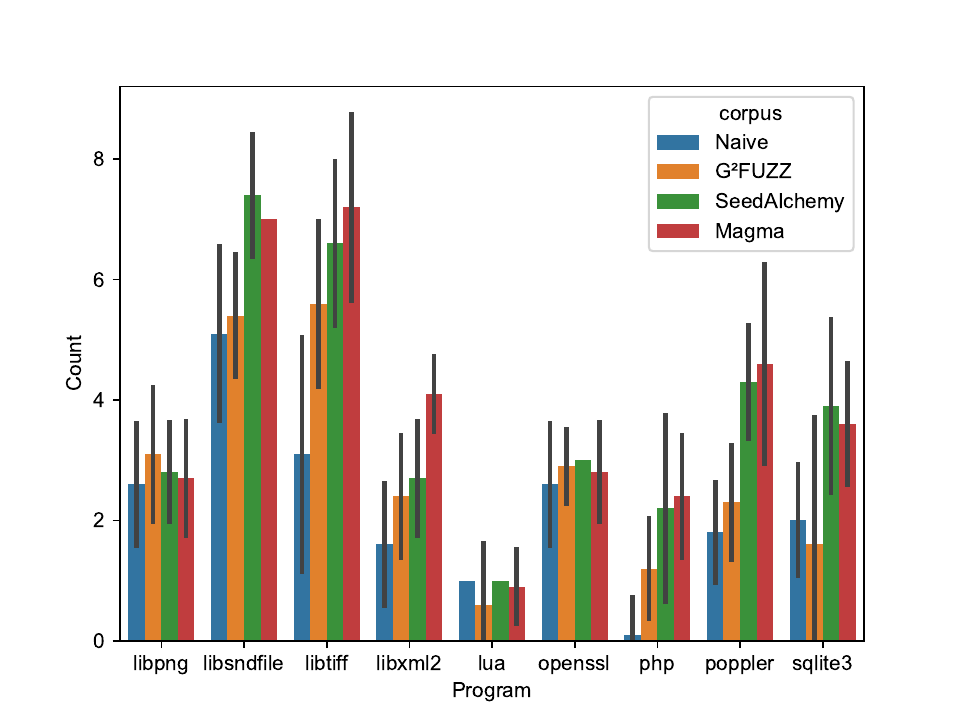}\label{fig:magma_triggered_bar_chart}}
    \hfil
    \subfloat[Code Coverage]{\includegraphics[width=\linewidth]{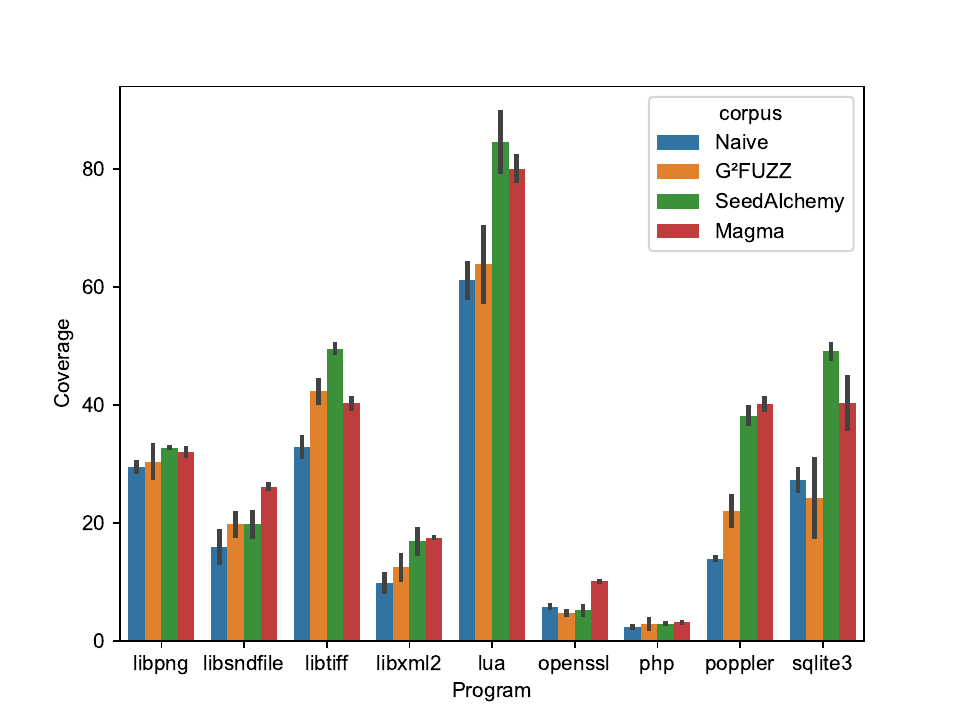}\label{fig:magma_coverage_bar_chart}}
    
    \caption{Bugs reached, bugs triggered, and code coverage for each target averaged over 10 trials. Error bars show the 95\% confidence interval.}
    \label{fig:magma_program_bar_charts}
\end{figure}

\figurename~\ref{fig:magma_reached_bar_chart} and \ref{fig:magma_triggered_bar_chart} show the fuzzing performance of each corpus separated by targets. SeedAIchemy reached more bugs in 78\% (7/9) of the targets and triggered more bugs in 89\% (8/9) of the targets compared to the Naive and G$^2$FUZZ corpora. Compared to the Magma corpus, SeedAIchemy had an equal or higher number of bugs reached and bugs triggered in over 50\% of the targets. Overall, SeedAIchemy performed similarly to the Magma corpus on all targets except for openssl and libxml2 in the number of bugs reached and bugs triggered respectively. 

\begin{figure}[tb]
    \centering
    \includegraphics[width=\linewidth]{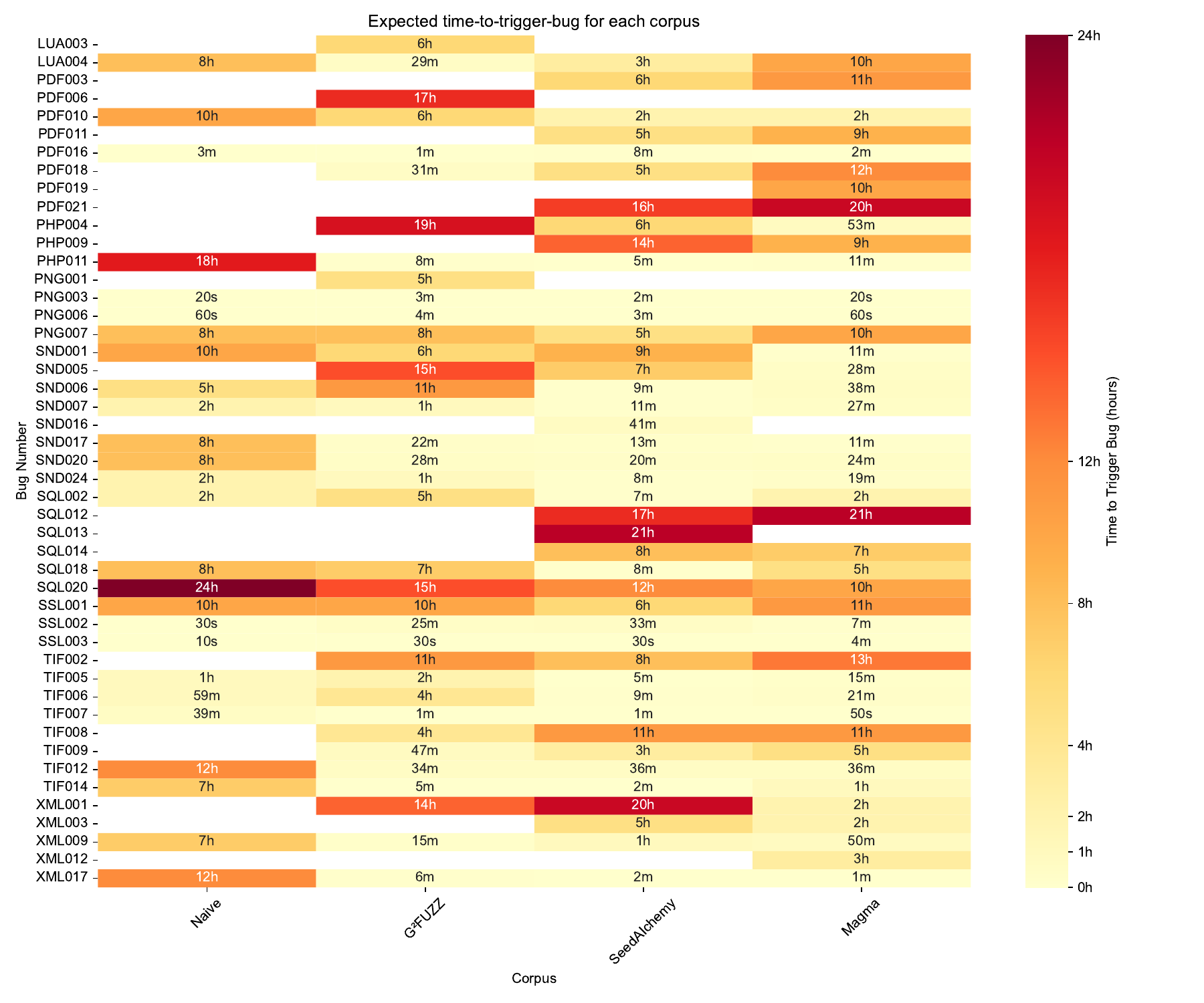}
    \caption{Time to trigger bug averaged across 10 trials.}
    \label{fig:bugs_triggered_heat_map}
\end{figure}

\begin{figure}[tb] % instead of figure*
    \centering
    \includegraphics[width=\linewidth]{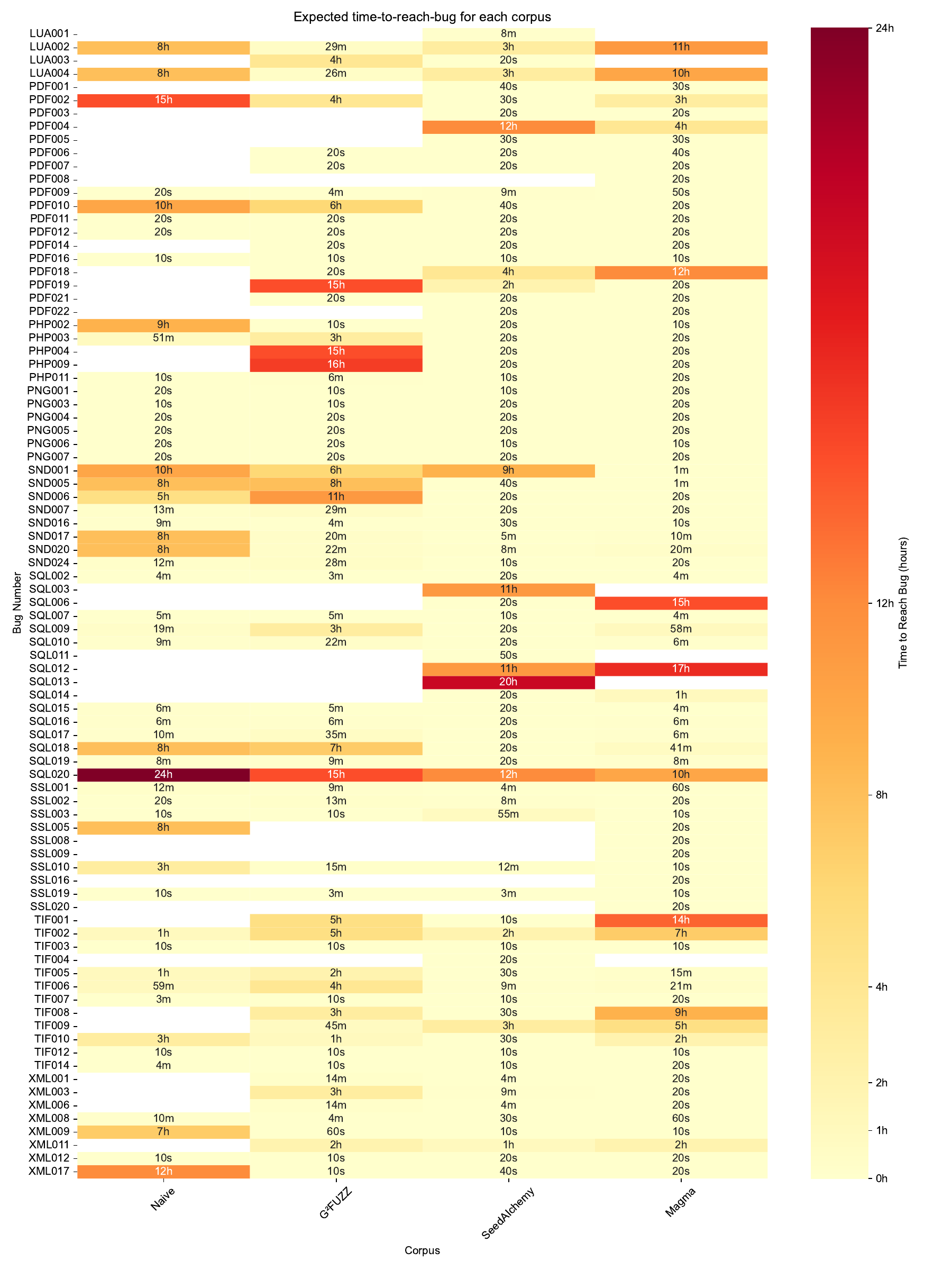}
    \caption{Time to reach bug averaged across 10 trials.}
    \label{fig:bugs_reached_heat_map}
\end{figure}

\figurename~\ref{fig:bugs_triggered_heat_map} and \ref{fig:bugs_reached_heat_map} show the average time to reach and time to trigger a bug for each corpus. For clarity of presentation, the reported times were rounded to the nearest second, minute, or hour, as appropriate. Across the set of bugs triggered by both the Magma and SeedAIchemy corpora, SeedAIchemy had the fastest triggering time in 55.0\% (22/40) of bugs. Across the set of bugs reached by both corpora, SeedAIchemy was fastest at reaching 52.0\% (39/75) of the bugs.

Based on Wilcoxon signed-rank tests\footnote{For each target, we computed the average number of bugs reached/triggered across 10 trials and paired these averages between corpora. Then, we applied the one-sided Wilcoxon signed-rank test to the pairs using the null hypothesis that $\text{median}(X) = \text{median}(Y)$, where $X$ and $Y$ denote the paired number of bugs reached/triggered of two corpora.}, SeedAIchemy significantly outperformed the Naive and G$^2$FUZZ corpora on bugs reached and triggered ($p < 0.01$ for all comparisons) while showing no significant difference from Magma corpora (reached: $p = 0.4253$; triggered: $p = 0.4687$). Table \ref{tab:wilcoxon} summarizes the Wilcoxon signed-rank test results comparing SeedAIchemy with other corpora. 

\begin{table}[tb]
\centering
\caption{Wilcoxon Signed-Rank Tests Comparing SeedAIchemy to Other Corpora}
\label{tab:wilcoxon}
\begin{tabular}{l l c}
\toprule
\textbf{Comparison} & \textbf{Bug Type} & \textbf{p-value} \\
\midrule
SeedAIchemy vs. Naive   & Reached   & 0.00116 \\
                        & Triggered & 0.00110 \\
SeedAIchemy vs. G$^{2}$FUZZ & Reached   & 0.00094 \\
                        & Triggered & 0.00789 \\
SeedAIchemy vs. Magma   & Reached   & 0.42525 \\
                        & Triggered & 0.46874 \\
\bottomrule
\end{tabular}
\end{table}

\subsection{Code coverage}

We evaluated the average code coverage for each corpus as shown in \figurename~\ref{fig:code_coverage_over_time} and \ref{fig:magma_coverage_bar_chart}. To bring coverage scores into a common range, we normalized the data by dividing each coverage data point by the final 24 hour coverage of the associated Magma corpus coverage value so that values higher than 1 represent better coverage than the Magma corpus and values lower than 1 represent worse coverage. Overall, code coverage with the SeedAIchemy corpus was significantly better than the naive and G$^2$FUZZ corpus, and not quite as good as the Magma corpus (\figurename~\ref{fig:code_coverage_over_time}).

%To follow suggested evaluation metrics by \cite{klees2018evaluating}, we also evaluate the effectiveness Figure \ref{fig:code_coverage_over_time} Figure \ref{fig:allcodecoverage} of SeedAIchemy against Magma and naive corpus using bitmap coverage over time generated by AFL++ \cite{aflplusplus-docs} as a supplementary measurement. The code coverage was plotted with similar methodology as bug coverage. Due to space limitations, we choose a subset of interesting plots to present as Figure \ref{fig:allcodecoverage} in the appendix. Shaded regions in the plots denote 95\% confidence intervals. To determine superiority of code coverage between one corpus and another, we declare a tie when the confidence intervals of two corpora overlap at the end of the experiment; otherwise, we consider the corpus with higher final mean code coverage the winner. 

Looking at individual targets in \figurename~\ref{fig:magma_coverage_bar_chart}, SeedAIchemy outperformed Magma's corpus on 44\% (4/9) of the targets
% , had similar performance on 33\% (3/9) of the targets, 
and underperformed on 56\% (5/9) of the targets.
% We categorize results as outperforming if a corpus’s mean coverage lies above the other’s 95\% CI, underperforming if below, and similar if within. 
Compared to the Naive and G$^2$FUZZ corpora, SeedAIchemy had higher or equivalent coverage on all targets except openssl, where the Naive corpus slightly outperformed the G$^2$FUZZ and SeedAIchemy corpora. 

%We also analyzed a combined metric \ref{coverage_over_time}\ref {coverage_over_time_most} which normalized the curves and averaged them across targets. We choose dividing by mean normalization for this case in order to have clearer comparisons between the curves, which could often have absolute values with large differences (.2-.8). The results showed the tool underperforming below the confidence interval of 95\%. The chart that excludes the targets with uncommon, unsupported types shows a smaller difference but still below the confidence interval.

\subsection{Discussion} 
% The naive corpus failed to trigger or reach any bugs on most of the target programs. After inspecting the fuzzing logs, we found that AFL++ often got stuck on the original corpus and wasn't able to generate new seed files. Thus, the performance of the naive corpus may be heavily influenced by the selection of the fuzzer. 

Our results show that SeedAIchemy generates high-quality corpora with fuzzing performance that is comparable to a manually constructed one. Corpora generated by SeedAIchemy significantly outperform Naive corpora and G$^2$FUZZ corpora in bugs reached,  bugs triggered, and code coverage. Our results demonstrate that SeedAIchemy is a suitable replacement for a manually constructed corpus on a wide variety of fuzz targets.

The fact that SeedAIchemy's corpora outperform those generated by G$^2$FUZZ suggests that fuzzing performance is better when LLMs are prompted to generate search queries for seed files rather than generate seed files directly. Searching for files using search queries allows SeedAIchemy to take advantage of the diversity of files on the internet, which is hard for an LLM to replicate on its own. SeedAIchemy's approach is also more cost effective because it only relies on a few LLM queries to produce a fully functioning seed corpus. By automating the process of seed corpus generation, SeedAIchemy reduces the time and effort it takes to construct a high quality seed corpus for fuzzing.

One potential concern in evaluating on benchmarks like Magma constructed from past bugs is the possibility of data leakage. Before fuzzing on the generated corpus, we filtered out any seeds that directly cause a crash to the target program, which reduces the risk of inadvertent leakage. Nevertheless, we acknowledge that it is difficult to rule out entirely such ``time-traveling'' advantage for SeedAIchemy, since any corpus generation method that relies on existing seed repositories could include posterior inputs that trigger prior bugs.

\section{Related Work}
% Having described our approach and results, we now turn to prior research to compare our contributions within the broader field of fuzzing. 
Fuzzing has emerged as one of the most effective automated techniques for discovering software vulnerabilities. It works by executing programs with large volumes of generated or mutated inputs and monitoring their behavior for crashes or anomalies \cite{manes2019art}, \cite{zhu2022fuzzing}. The quality of the seed corpus is a critical factor in fuzzing effectiveness \cite{herrera2021seed}, \cite{rebert2014optimizing}, \cite{wang2017skyfire}. 

\subsection{Corpus Generation and Optimization}
Traditionally, fuzzing has relied on domain experts manually collecting representative inputs \cite{padhye2019semantic}. Some studies evaluate fuzzing with  a single seed file, which measures the effectiveness of fuzzing when little attention is put into corpus construction.  Other studies use a corpus of files collected by fuzzing teams or crawling public repositories and the web \cite{godefroid2017learn}. 

Constructing a comprehensive corpus remains challenging and time-consuming for developers \cite{li2018fuzzing}, \cite{wang2020systematic}. Our work automates corpus construction by finding existing corpora from open-source projects and supplementing them with a wide variety of seed files scraped from the web. We apply corpus minimization, which improves fuzzing efficiency by removing redundant seed files from the corpus \cite{herrera2021seed}. With this approach, we aim to lower the barrier to corpus generation \cite{nourry2023human}, while improving the overall effectiveness of fuzzing.

\subsection{AI in Corpus Generation}
AI and machine learning have been previously applied to corpus and seed generation to address the limitations of manual approaches. Studies have used generative models to automatically produce syntactically valid C programs for fuzzing compilers \cite{liu2019deepfuzz} or augment corpora \cite{nichols2017faster}, \cite{lyu2018smartseed}.  However, these approaches are typically limited in their ability to support binary file formats and long-tail less widespread file formats.  

Recent studies have focused on leveraging LLMs for seed corpus generation. \cite{Xu2024ISC4DGF} uses LLMs to generate and select corpora tailored to known vulnerabilities; \cite{shi2024harnessing} uses LLMs to generate a script that produces seeds and iteratively improves upon them based on the code coverage data; similarly, \cite{Zhang2025G2FUZZ} uses LLMs to write code to generate seed files, making use of existing libraries for these file formats; \cite{AutoCorpus2023} is an open source tool that uses LLMs to generate seed files for text files in widely used formats like JSON or XML; \cite{Asmita2024BusyBox} uses LLMs for target-specific fuzzing seed generation to identify vulnerabilities efficiently.

We explore a different approach: rather than using AI to generate valid input files, we collect input files from the web by using AI to construct intelligent web search terms that help us find a diverse set of seed files. Our strategy is less computationally intensive than training generative models, remains portable across domains without retraining, works for non-textual file formats, and allows practitioners to inspect and refine the generated queries for greater transparency and control \cite{saavedra2019review}.

% \subsubsection{Evaluation and Benchmarking}
% We follow robust evaluation and benchmarking principles proposed in prior work \cite{klees2018evaluating}, \cite{schloegel2024sok}, \cite{bohme2022reliability}. For example, we conduct 10 independent trials, fuzz for 24 hours, and count the number of unique vulnerabilities (not just coverage).

\section{Future Work}
% - Performs poorly on uncommon formats that lack extensions or formats with very specific data structures that aren't governed by filetype
% - Review next steps/what we could have better?

Although SeedAIchemy improves the process of generating seed corpora, there are some areas for future improvement:
\begin{itemize}
    % \item \textbf{Handling seed files without file extensions:} SeedAIchemy cannot handle rare or ambiguous filetypes, such as raw bytecode, proprietary encoding, or formats without file extensions. Future work could complement the existing tool methodology by adding support for custom file types.
    % \item \textbf{Parallelization}: Due to rate limits of the Google Cloud Custom Search API, SeedAIchemy cannot generate multiple corpora in parallel. Thus, fuzz targets that take multiple file types as input require longer corpus generation times. Future work could investigate how to improve the efficiency of SeedAIchemy so that it can be run in parallel. 
    \item \textbf{Testing multiple fuzzers}: Our evaluation was conducted exclusively with AFL++, which represents one of many widely used fuzzers. Future work could investigate the effectiveness of SeedAIchemy across different fuzzers such as libFuzzer \cite{LLVMlibFuzzer} or Honggfuzz \cite{Honggfuzz}.
    \item \textbf{Testing individual module contributions}: We have not evaluated the contributions of each individual module on the performance of the overall corpus. Future work could investigate whether certain modules contribute more to bug discovery and code coverage than others. 
\end{itemize}

\section{Conclusion}

% Bullet points/outline: 
% - We showed that an LLM-guided can achieve similar results to a curated corpus on well-typed formats. 
% - Performs poorly on uncommon formats that lack extensions or formats with very specific data structures that aren't governed by filetype
% - Final results: [fill in later] based on amount of bugs hit, speed, coverage? any other metrics?
% - SeedAIchemy is best used when users is fuzzing a program with file structure defined by the file type; user also should be willing to minimize the corpus for best results
% - Review next steps/what we could have better?

In this paper, we present SeedAIchemy, an LLM-driven seed corpus generation tool for fuzzing. The corpora produced by SeedAIchemy perform significantly better than naive corpora and similarly to expert-curated corpora. SeedAIchemy also outperforms corpora generated by G$^2$FUZZ, an existing LLM-based corpus generation tool that prompts LLMs to generate seed files directly. SeedAIchemy makes the fuzzing process cheaper and more accessible to developers who are trying to incorporate fuzzing into their development cycle by providing an automated tool for generating high-quality corpora.

\section*{Acknowledgment}
\addcontentsline{toc}{section}{Acknowledgment}
This work was supported by the  KACST-UC Berkeley Center of Excellence for Secure Computing and the NSF ACTION center through NSF grant 2229876. 

\bibliographystyle{IEEEtran}
\bibliography{references.bib}

\clearpage

\begingroup

\onecolumn
\appendix

\begin{figure*}[h!]
    \centering
    \includegraphics[scale=0.32]{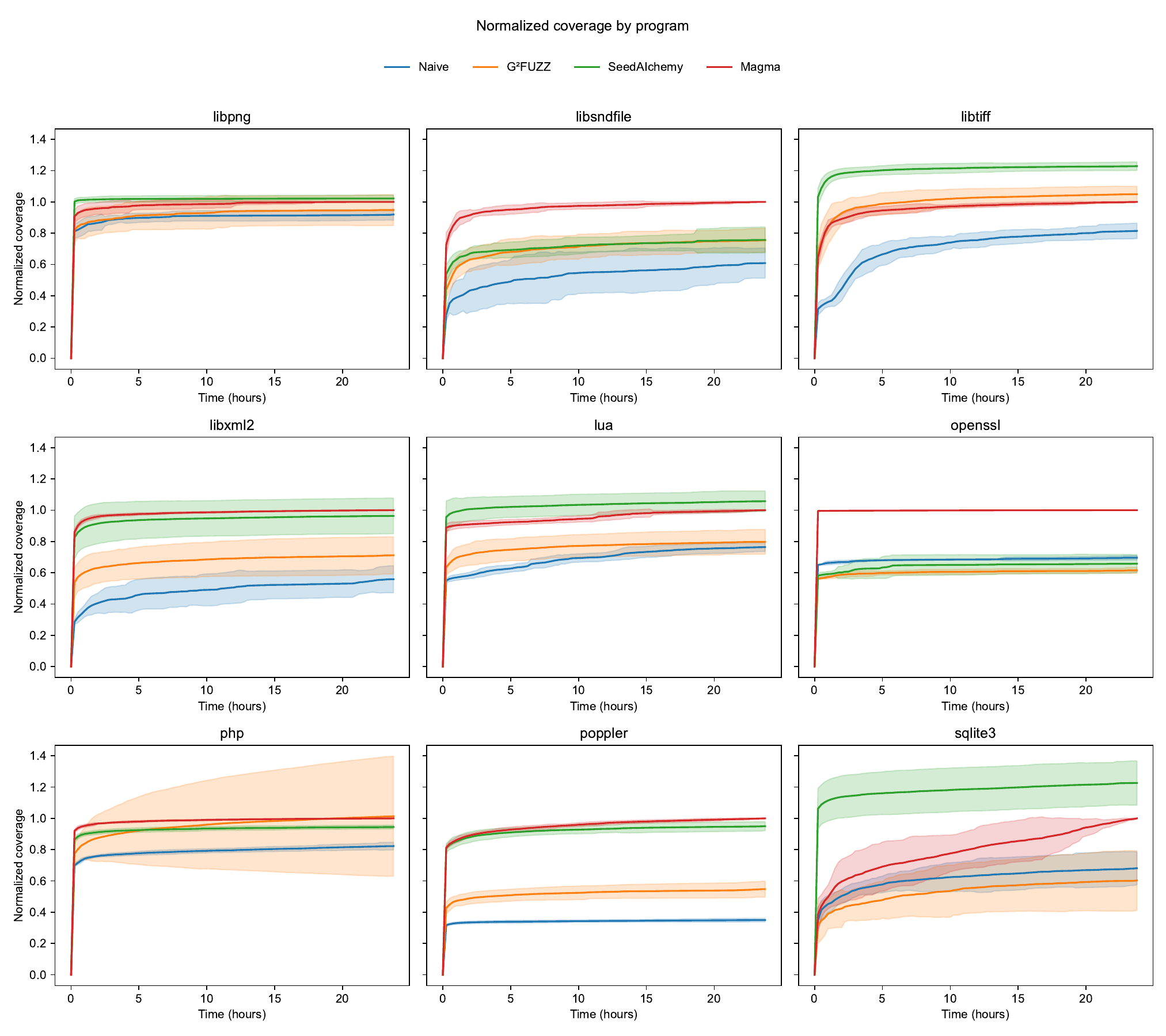}
    \caption{Normalized coverage for each Magma target}
\end{figure*}

\begin{figure*}[h!]
    \centering
    \includegraphics[scale=0.32]{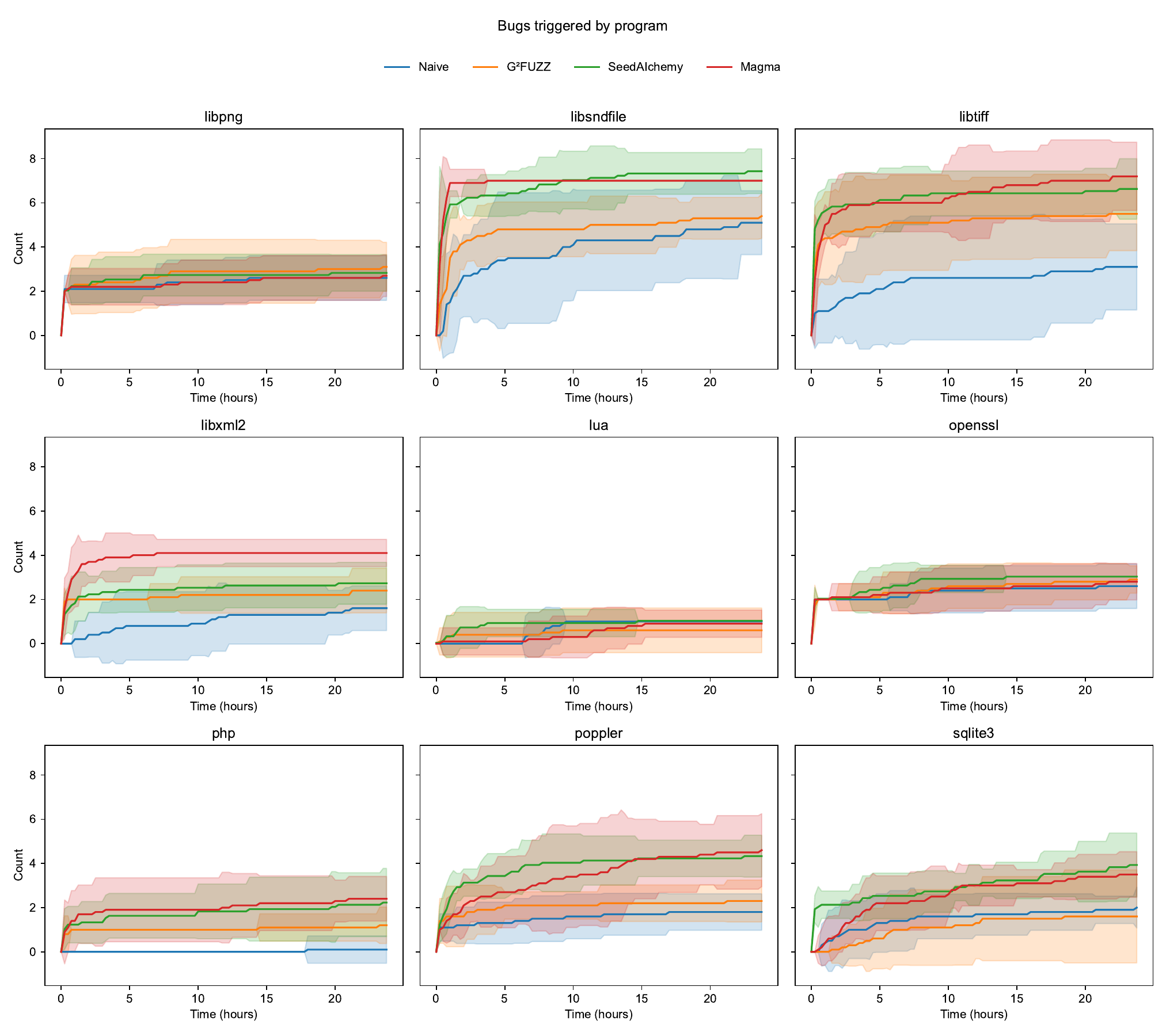}
    \caption{Bugs triggered for each Magma target}
\end{figure*}

\endgroup

\end{document}